\documentclass[prl,preprintnumbers,showpacs,twocolumn]{revtex4}

\usepackage{graphicx}
\usepackage{bm}

\begin{document}
\preprint{}
   \title{Response of Complex Systems to Complex Perturbations: \\Complexity Matching}

\author{Paolo Allegrini$^{1}$}
\author{Mauro Bologna$^{2}$}
\author{Paolo Grigolini$^{3,4,5}$}
\author{Mirko Lukovic$^{4}$}

\affiliation{$^{1}$Physics Department, Duke University, Durham NC 27708 USA.\\
$^2$Departamento de F\'{i}sica,
Universidad de Tarapac\'{a}, Campus Vel\'{a}squez, Vel\'{a}squez
1775, Casilla 7-D, Arica, Chile.\\
$^3$Center for Nonlinear Science, University of North Texas,
    P.O.Box 311427, Denton TX 76203-1427 USA.\\
$^{4}$Dipartimento di Fisica "E. Fermi"-Universit\`a di Pisa and
INFM, Largo Pontecorvo 3, 56127 Pisa, Italy.\\
$^{5}$Istituto dei Processi Chimico Fisici del CNR
Area della Ricerca di Pisa, Via G. Moruzzi 1,
56124 Pisa, Italy.}
\date{\today}

\begin{abstract}  We argue that complex systems, defined as non-Poisson renewal process, with complexity index $\mu$, exchange information through complexity matching. We illustrate this property with detailed theoretical and numerical calculations describing a system with complexity index $\mu_{S}$ perturbed by a signal with complexity index $\mu_{P}$. We focus our attention on the case $1.5 \leq \mu_S \leq 2$ and $1 \leq \mu_{P} \leq 2$. We show that for $\mu_{S} \geq \mu_P$, the system S reproduces the perturbation, and the response intensity increases with increasing $\mu_P$. The maximum intensity is realized by the matching condition $\mu_P = \mu_S$. 
For $\mu_{P} > \mu_{S}$ the response intensity dies out as $1/t^{\mu_P-\mu_S}$.
 \end{abstract}

\maketitle
 
Aperiodic stochastic resonance \cite{aperiodic} and phase synchronization \cite{phase} are recently discovered phenomena through which physicists try to establish a control on complex processes, notably those of neuro physiological interest. 
We study an intermittent behavior, reminiscent of the noise-free stochastic resonance process through chaotic maps \cite{chaoticmap1,chaoticmap2}. However, we go beyond the choice of a driving force either of Poisson kind or totally random. 
As to the synchronization, we establish it using the concept of Complexity Matching (CM). 
Our departure point is the recent observation \cite{recentobservation,barbi} that a non-Poisson renewal system with power index $\mu_{S} < 2$ does not respond to a  harmonic perturbation. We devote this letter to proving that a non-Poisson renewal system with index $\mu_S < 2$, on the contrary, is so sensitive to non-Poisson renewal perturbations with power index $\mu_{P} <2$ as to inherit in the long-time limit their power indexes.  Furthermore, we shall show that the response intensity is maximum 
at the matching condition $\mu_S = \mu_{P}$, and that, when departing from it, the response intensity either will be weaker $(\mu_S > \mu_P)$ or will die out $(\mu_S < \mu_P)$.

The CM phenomenon is a property of  Non-Poisson Renewal (NPR) processes, here considered as a  paradigmatic example of complexity. The  Blinking Quantum Dots (BQD) \cite{bqd} are NPR systems \cite{brokman}: The jump from the ``on" (``off") state to the ``off" (``on") resets to zero the system's memory, thereby ensuring the renewal condition. Furthermore, the time distance between two consecutive events is described by a histogram with the form of an inverse power law with power index $\mu_{S} < 2$. The free dynamics of the process are described by the dichotomous variable $\xi_{S}(t)$, with the values $\xi_S = 1$ and $\xi_S = -1$ indicating that the system is in the ``on" (or $|1\rangle$) and ``off" (or $|2\rangle$) state, respectively. 
As pointed out by Bel and Barkai \cite{bel}, the condition $\mu_{S} < 2$ of the NPR systems generates ergodicity breakdown, a condition shared by the phenomenological models of glassy dynamics \cite{glassy}, laser cooling \cite{book} and models of atomic transport in optical lattices \cite{atomic}.

To ease the theoretical treatment, we make the simplifying assumption that the ``on" waiting time distribution is identical to the ``off" waiting time distribution. When the system moves from an initial out of equilibrium condition, the survival probability $\Psi_{S}(t)$ is given by 
\begin{equation}
\Psi_{S} (t) = \left [\frac{T_S}{t + T_S} \right ]^{\mu_S-1}.
\end{equation}
This experimentally observable relaxation is traced back \cite{gerardo} to the waiting time distribution
\begin{equation}
\label{theoretical}
\psi_S(\tau) = (\mu_S -1) \frac{T_{S}^{\mu_S - 1}}{(\tau + T_S)^{\mu_S}}.
\end{equation}
This distribution refers to the distance between two consecutive collisions, which are as renewal as jumps, but are not necessarily jumps. At the collision occurrence the system has to decide, through a fair coin tossing prescription, whether to jump to the other state or to remain in the original state. 
This distribution is properly normalized, and the parameter
$T_S$, making this normalization possible, gives information on the lapse of
time necessary to reach the time asymptotic condition at which $\psi_S(\tau)$
becomes identical to an inverse power law. 

To express the effect of a perturbation on the system, we limit ourselves to assuming that 
\begin{equation}
\label{external}
T^{(\pm)}_S(t) = T_S(0)(1\pm\epsilon \xi_{P}(t)), 
\end{equation}
where $\epsilon$ is the perturbation strength and $\xi_{P}(t)$ denotes the perturbation signal. A concise account of the dynamic arguments of Refs. \cite{barbi,gianluca} is as follows. In the limiting case $\epsilon \rightarrow 0$, a collision occurring in the ``on" (``off") state at time $t^{\prime}$, earlier than the collision occurring at $t > t^{\prime}$ in the ``off" (``on") state, generates  the response function given by the $t^{\prime}$-aged waiting time distribution 
$\psi_{S}(t,t^{\prime})$: This is 
the probability density for the unperturbed system, prepared at $t=0$, to undergo a collision
at time $t$, with no collisions happening in the interval $(t^{\prime},t)$, given that observation begins at time $t^{\prime}$.
Thus, we have \cite{barbi,gianluca}
 \begin{equation}
 \label{key}
\Pi(t) \equiv  \langle \xi_{S}(t) \rangle =  \epsilon \int_{0}^{t} dt^{\prime} \psi_{S}(t,t^{\prime}) \xi_{P}(t^{\prime}).
 \end{equation}
  Here $\Pi(t) \equiv p_{1}(t) - p_{2}(t)$, with $p_{1}(t)$  and $p_{2}(t)$ denoting the probability that at time $t$ the system is found either in the state $|1\rangle$ or in the state $|2\rangle$.

 Let us explain now how to realize the complex perturbation $\xi_{P}$. The simplest possible way is to adopt as a perturbation the signal $\xi_P(t)$ produced by another two-state NPR system.
 The single realization $\xi_{P}(t)$ is the perturbation signal, to which the system $S$ has to respond. Let us prove first that the perturbed signal $\xi_{S}(t)$ inherits the complexity index $\mu_P$. Let us assume that we have at our disposal an ensemble of perturbation signals $\xi_{P}(t)$, all of them properly prepared at time $t=0$, and let us study the double average $\langle\langle\xi_{S}(t)\rangle\rangle$. We realize the perturbation preparation by 
 setting all the perturbation signals at the beginning of their sojourn in the ``on" state. 
 It is known \cite{gerardo} that in this case \begin{equation}
\label{thesame}
 \langle\xi_{P}(t)\rangle = \Psi_{P} (t) = \left (\frac{T_P}{t + T_P} \right )^{\mu_P -1}.
 \end{equation} 
 Thanks to the linear nature of Eq. (\ref{key})
 we obtain
 \begin{equation}
 \label{key22}
\langle\Pi(t)\rangle \equiv  \langle\langle\xi_{S}(t)\rangle\rangle =  \epsilon \int_{0}^{t} dt^{\prime} \psi_{S}(t,t^{\prime}) \Psi_{P}(t^{\prime}).
\end{equation}
We establish the time asymptotic properties of $\langle \Pi(t) \rangle$, Eq. (\ref{key22}), by adopting the Fourier expansion of $\Psi(t) = \int d\omega \xi_{P}(\omega) exp(-i\omega t)/(2\pi)$, which allows us to use the response to a monochromatic perturbation of Ref. \cite{barbi, gianluca}: Using this result we obtain that for $t \rightarrow \infty$,
\begin{equation}
\langle\Pi(t)\rangle \propto \frac{\epsilon}{\Gamma(\mu-1)}\int d\omega \xi_{P}(\omega) \frac{e^{-i(\omega t + \frac{\pi \mu}{2})}}{(\omega t)^{2-\mu_S}} .
\end{equation}
We note that for $\omega \rightarrow 0$ $\xi_{P}(\omega) \propto \omega^{\mu_{P}-2}$. Thus, it is straightforward to prove that for $t\rightarrow \infty$ $\langle\Pi(t)\rangle\propto1/t^{\mu_{P}-1}$, with the condition
\begin{equation}\label{nodivergency}
\mu_S + \mu_P > 3,
\end{equation} 
whose violation would make the integration over $\omega$ divergent. Moving along these lines we explore the ranges $1.5 < \mu_{S} < 2$ and $3-\mu_S < \mu_{P} < 2$, fitting the requirement of Eq. (\ref{nodivergency}), and we show
\begin{equation}
 \label{key3}
\frac{ \langle\Pi(t)\rangle}{\epsilon}= \frac{k_{1}(\mu_{S},\mu_{P})}{t^{\mu_{P} -1}}  + \frac{k_{2} (\mu_{S},\mu_{P})}{t^{\mu_{P} + 1 -\mu_{S}}},
 \end{equation}
 where
 \begin{equation}
 \label{responseintensity}
 k_{1}(\mu_S,\mu_P) = -\frac{ \sin\pi \mu_S}{\pi} \frac{\Gamma (2 - \mu_S)\Gamma(1-\mu_P + \mu_S)}
 {T_{P}^{1-\mu_P }\Gamma(3-\mu_P)}.
 \end{equation}
 For concision's sake we do not write the somewhat extended explicit expression of $k_{2}(\mu_S,\mu_P)$, which is not used by the heuristic theoretical prescriptions of this letter, the only  property worth of mention being that 
$k_{2}(\mu_S,\mu_P)>0$, if $\mu_{S}<\mu_P$ and
$k_{2}(\mu_S,\mu_P) < 0$, 
 if $\mu_{S}>\mu_P$.
 These analytical and exact results prove that the system inherits the complexity index $\mu_P$ of the perturbing signal. 
 
 In the case $\mu_P < \mu_S$  it is possible to establish a reasonable connection between the statistical mean of Eq. (\ref{key22}), expressed in the asymptotic time limit by Eq. (\ref{key3}),  and the response to a single realization of perturbing signal.
 If the system is fast enough as to adapt itself to the perturbation signal $\xi_{P}$, as it happens in the case of ordinary stochastic resonance
 \cite{aperiodic,stochasticresonance}, the mean response should reproduce the perturbation signal $\xi_{P}$. We make the more cautious assumption that the mean average adapts itself to the signal $\xi_{P/S}(t)$, which is is very close to $\xi_{P}(t)$, and fits the requirement
 \begin{equation}
 \label{weakrequirement}
\langle\xi_{P/S}(t)\rangle = \langle\xi_{P}(t)\rangle = \Psi_{P}(t). 
 \end{equation}
 Thus, we make the heuristic proposal 
  \begin{equation}
 \label{heuristic}
 \Pi(t)=\epsilon [1 - \pi_{S}(t)] A(\mu_{S},\mu_{P}) \xi_{P/S}(t).
\end{equation}
The function $\pi_{S}(t)$ denotes the learning function, whose time asymptotic property, on the basis of the response to constant perturbation \cite{barbi,gianluca}, is assumed to be
 \begin{equation}
 \pi_{S}(t) \propto {t^{\mu_S-2}}.
 \end{equation}
 Using Eq. (\ref{weakrequirement}), the average over the single realizations of Eq. (\ref{heuristic}) yields
 \begin{equation}
 \label{heuristic_mean}
 \langle \Pi(t)\rangle=\epsilon[1 - \pi_{S}(t)] A(\mu_{S},\mu_{P}) \Psi_P (t),
\end{equation}
which is compatible with the theoretical prescription of Eq. (\ref{key3}). Thus, using Eq. (\ref{responseintensity}) and the asymptotic time expression for Eq. (\ref{thesame}), we conclude that for $\mu_{S} > \mu_P$ the amplitude
$A$ is given by $A(\mu_S,\mu_P)  = k_{1}(\mu_S,\mu_P)/T_P^{\mu_P-1}$.  
 
  We now have recourse to a numerical treatment for the twofold purpose, $(i)$, of establishing the whole time evolution of $\langle \Pi(t) \rangle$ from $t=0$ to  $t=\infty$, and not only the asymptotic time behavior of Eq. (\ref{key3}), and, $(ii)$, of establishing the single realizations $\Pi(t)$. 
 We consider a coordinate $y$ moving within the interval $I = (0,1]$ with the equation of motion 
  \begin{equation}
  \label{stochastic}
  \dot{y} = a_{\pm}(t) y^{z_{S}},
  \end{equation}
  where $z_S \equiv \mu_S/(\mu_S - 1)$ and
  $a_{\pm}(t) = (\mu_S - 1)/T^{\pm}_{S}(t)$. The motion of $y$ within the interval $I$ corresponds to the system either in the ``on" or ``off" state. When the coordinate $y$ reaches the border value $y=1$ a new initial condition within the interval $I$ is randomly chosen with uniform probability. Then we toss a coin to decide whether to adopt $a_{+}(t)$ or $a_{-}(t)$ for the next phase of this dynamic process. This is the dynamic prescription used in \cite{gianluca} to yield the time dependence of $T_{S}$, Eq. (\ref{external}).

\begin{figure}[h]
\begin{center}
\includegraphics[scale = 0.35]{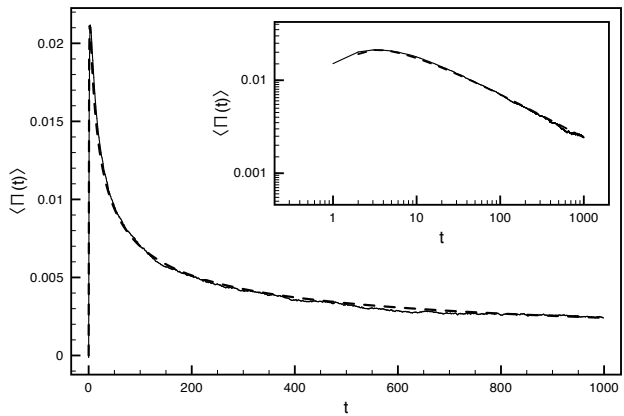}
\end{center}\vspace{-.8cm}
\caption{Response $\langle \Pi(t)\rangle$ for $\mu_P=1.5$ ($T_P=1$) and $\mu_S=1.6$ ($10^8$ systems with $T_S =1$, coupling $\epsilon=0.1$). 
We assign to the learning function the form $\pi_S(t)=B^{2-\mu_S}/(B+t)^{2-\mu_S}$. To fit the data (solid line) we assign to the fitting parameters $A$ and $B$ of the overall fitting function (\protect\ref{heuristic_mean}) (dashed line) the values $A = 0.80$ (in agreement with $k_1(1.6,1.5)=0.798$) and $B = 0.566$.
Insert: the same in logarithmic scale.}
\label{fit}
\end{figure}

\begin{figure}[h]
\begin{center}
\includegraphics[scale = .8]{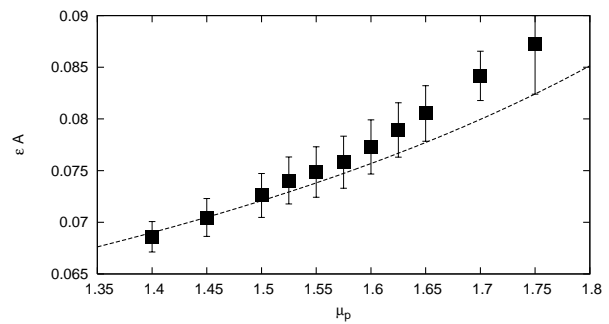}
\end{center}\vspace {-.3cm}
\caption{
The amplitude of $\epsilon A$ as a function of $\mu_P$, with $T_P = 1$, $\mu_S = 1.6$, $T_S = 1$, and $\epsilon = 0.1$ (average over $10^8$ systems). The dots indicate the result of the numerical fitting procedure illustrated in the text. For $\mu_P < \mu_S$ the procedure rests on Eq. (\protect\ref{heuristic_mean}). For $\mu_P \geq \mu_S$ the fitting function adopted is $A/t^{mu_P-1}$, with $ t > 100$. 
The solid line is the theoretical 
$\epsilon k_1(1.6,\mu_P)$.
}
\label{A-k1}
\end{figure}

  Note that Eq.  (\ref{key22}) is equivalent to
 \begin{equation}
 \label{key222}
 \Pi(t) = \epsilon \int_{0}^{t} dt^{\prime} \psi_{S}(t,t^{\prime}) \xi_{P}(t^{\prime}), \end{equation}
 with
 \begin{equation}
 \label{perturbationsignal}
 \xi_{P}(t) = \left (\frac{T_P}{t + T_P} \right )^{\mu_P - 1}.
 \end{equation}
Thus, the single trajectories $\xi_{S}(t)$ are determined by running Eq. (\ref{stochastic}) with the time dependent prescription of Eq. (\ref{external}), the proper random back injection and the choice of Eq. (\ref{perturbationsignal}) for the perturbation signal. 
 This allows us to check the heuristic prescription of Eqs. (\ref{heuristic}) and (\ref{heuristic_mean}). In Fig. \ref{fit} we show $\langle \Pi(t)\rangle$ corresponding to $\mu_{S} = 1.6$ and $\mu_P = 1.5$. We see that Eq. (\ref{heuristic_mean}) affords a good fitting of this condition. Notice that the adoption of this fitting prescription allows us to determine $A$ and compare it with $k_1$. In Fig. \ref{A-k1} we compare the numerical value of $A$ to the theoretical prediction of Eq. (\ref{key3}). We see that the agreement between numerical and theoretical result is compatible with the statistical accuracy of the numerical treatment and it becomes worse when $\mu_{P} > \mu_{S}$, as a consequence 
 of $k_{2}$ becoming positive and so contributing to the slowly decaying tails. 


\begin{figure}[h]
\begin{center}
\includegraphics[scale = .8]{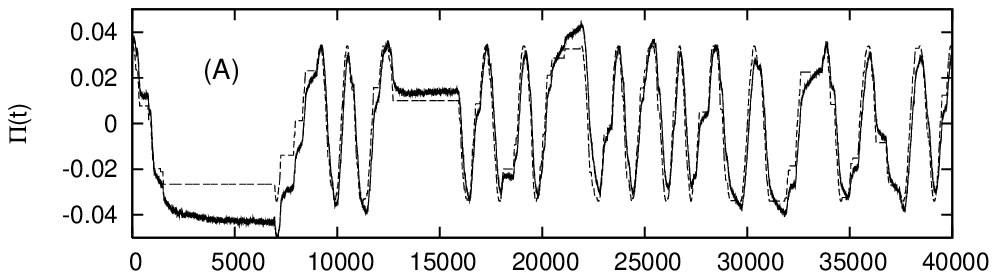}
\includegraphics[scale = .8]{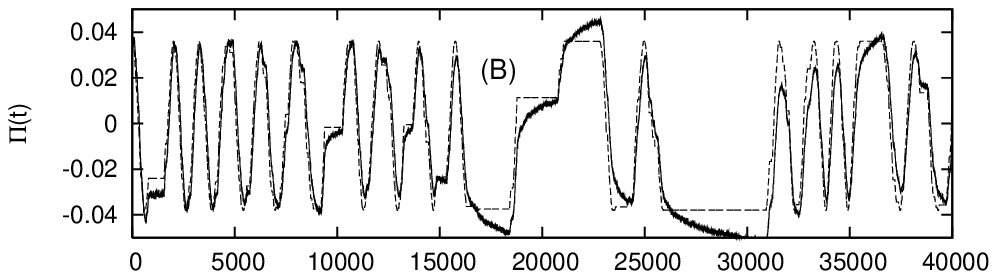}
\includegraphics[scale = .8]{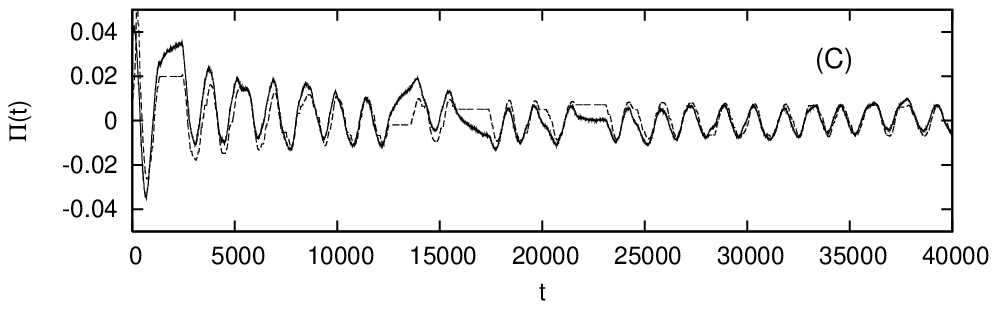}
\end{center}
\vspace {-.5cm}
\caption{
(A):  $\Pi(t)$ for $\mu_P=1.8$,  $\mu_S=1.9$. 
The two lines are the numerical response (solid) and, according to the modified heuristic prescription (\protect\ref{heuristic}), the signal $\xi_P$ multipied by $\epsilon A$ with $A = 0.34$ (dashed).
(B): Same for $\mu_P=1.9$, $\mu_S=1.9$. Here the fitting constant $A=0.37$.
(C):  $\mu_P=1.9$, $\mu_S=1.55$. 
The two lines are the numerical response (solid) and, according to the modified heuristic prescription (\protect\ref{first}), the signal $\xi_P$ multipied by $0.39\epsilon(\omega t)^{\mu_S-\mu_P}$ (dashed),
where $\omega=0.01$ is the frequency of the natural time harmonic function (\protect\ref{subord}).
}
\label{singlePI}
\end{figure}

 As to the single realizations, we have to create proper perturbation signals, which are complex and do not have abrupt jumps that may conflict with the assumptions made in Ref. \cite{barbi,gianluca} to derive the linear response theory of Eq. (\ref{key}). Thus, we  assume that in the natural time scale the perturbation is
  \begin{equation}
  \label{subord}
  \xi_{P}(n) = \cos (\omega n).
 \end{equation}
 To make this regular perturbation become complex we set first the condition $\omega \ll1$.
 Then, we turn the natural time $n$ \cite{subordination} into the continuous time $t$ with the prescription $t(n+1) = t(n) + \tau_{n}$, where $\tau_{n}$ is a number drawn from the distribution
 \begin{equation}
\label{theoretical2}
\psi_P(\tau) = (\mu_P -1) \frac{T_{P}^{\mu_P - 1}}{(\tau + T_P)^{\mu_P}},
\end{equation}
with $\xi_{P}(t) = \cos(\omega t_{n})$ for $t_{n+1} > t \geq t_n$. This procedure is inspired to an approach currently used \cite{subordination} to derive, for instance subdiffusion, by subordinating it, along the lines of Continuous Time Random Walk to ordinary diffusion. 
It is straightforward to prove that in the time asymptotic limit $\langle\xi_P(t)\rangle$ has the same power law as Eq. (\ref{thesame}). We follow the directions of Refs. \cite{barbi,gianluca} and we create many perturbed signals $\xi_{S}(t)$ by means of the prescription of Eq. (\ref{external}). We make the ensemble average on a very large number of responses ($2\cdot 10^6$) and we get the results illustrated in Fig. \ref{singlePI}.

The numerical results of Fig. \ref{singlePI} make it possible for us to make the heuristic prediction for the response $\Pi(t)$
in the case $\mu_S < \mu_P$,
  \begin{equation}
 \label{first}
 \Pi(t)=\epsilon\xi_{S/P}(t)  R(t)  , 
 \end{equation}
 where 
 \begin{equation}
 \label{second}
 R(t) \simeq k_{1}(\mu_S,\mu_P) {(\omega t)^{\mu_S-\mu_P}}.
 \end{equation}
 The signal $\xi_{S/P}(t)$ is again very similar to the perturbation $\xi_P(t)$. Also in this case a fluctuating departure of $\xi_{S/P}(t)$ from $\xi_P(t)$ is admitted. This fluctuation can be stronger than the fluctuation of $\xi_{P/S}(t)$ from $\xi_{P}(t)$. In this case we make the request
 \begin{equation}
 \label{weakrequirement2}
 \langle\xi_{S/P}(t) \rangle = \langle\xi_{S}(t) \rangle = \Psi_{S}(t) = \left(\frac{T_S}{t+T_S}\right)^{\mu_S-1}. 
 \end{equation}
 Thus, averaging Eq. (\ref{first}), and using Eqs. (\ref{first}) and (\ref{weakrequirement2}), we recover the same time asymptotic result as Eq. (\ref{key3}), and the constraint $\langle\Pi(t)\rangle \propto 1/t^{\mu_P-1}$ for $t\rightarrow \infty$.

  Fig. \ref{singlePI} refers to the condition $(\omega <<1)$ creating the same asymptotic behavior as $\Psi(t)$ of Eq. (\ref{thesame}) after an extended  intermediate asymptotic regime, thereby preventing us from adopting a quantitative comparison with  Eq. (\ref{key3}). However,  in a qualitative agreement with Eq. (\ref{key3}), Fig. \ref{singlePI} shows that for $\mu_S > \mu_P$ the response intensity is weaker than at the matching condition, and that for $\mu_{P} > \mu_{S}$ the response dies out. 
  This proves the CM effect, going beyond the theoretical results of Eq. (\ref{key3}) that might generate the wrong impression that the response intensity increases monotonically from $\mu_{P} = 1.5$ up to $\mu_{P} = 2$. It is worth pointing out that due to the numerical results of Figs. \ref{A-k1} and \ref{singlePI} as well as to the exact asymptotic time prediction of Eq. (\ref{key3}), for $\mu_P>\mu_S$ we have to replace the heuristic prescription of Eq. (\ref{heuristic}) with the heuristic prescription of Eq. (\ref{first}), which is, in a sense, equivalent to the response intensity falling to zero. Note that we have modified the heuristic prescriptions of Eqs. (\ref{heuristic}) and (\ref{first}) by replacing both $\xi_{S/P}(t)$ in the former case and $\xi_{P/S}(t)$ in the latter case, with $\xi_{P}(t)$.  
   The agreement between the numerical results and these modified heuristic prescriptions is remarkably good.

In conclusion, this letter shows that a complex system with $\mu_S < 2$ is sensitive to complex perturbations with $\mu_P <2$, so that the mean value $\langle\Pi(t)\rangle$ inherits the perturbation power index $\mu_P$. The maximum response is realized through the matching condition $\mu_{S} = \mu_P$. If $\mu_{P} < \mu_{S}$ the response intensity is given by $k_1(\mu_s,\mu_P)/T_P^{\mu_P-1}$ and gets the maximum value when the matching condition is realized. When $\mu_{P} > \mu_{S}$ the response intensity dies out. There are good reasons to believe that with $\mu_{S} > \mu_{P}$ the function $\xi_{P/S}(t)$ becomes identical to $\xi_P(t)$ thereby yielding a sort of generalization of the well known phenomenon of stochastic resonance \cite{aperiodic,stochasticresonance}. 

\emph{Acknowledgments} We 
thankfully 
acknowledge Welch foundation for partial support 
through Grant No. 70525.

\end{document}